%% file: WHRT-L2-performance.tex
\documentclass[10 pt, conference, letterpaper]{ieeeconf}

\IEEEoverridecommandlockouts                              %
\usepackage{amsmath,amssymb,amsfonts}
\usepackage{mathtools}
\usepackage{tikz}
\usepackage{pgfplots}
\let\labelindent\relax %
\usepackage{enumitem}
\usepackage[group-separator={,}]{siunitx}

\newcommand{\ie}{i.e., }
\newcommand{\eg}{e.g., }
\newcommand{\cf}{cf.\ }

\newcommand{\anyWHRTC}[2]{\genfrac{(}{)}{0pt}{}{#1}{#2}}
\newcommand{\rowWHRTC}[2]{\genfrac{\langle}{\rangle}{0pt}{}{#1}{#2}}
\newcommand{\diagdots}[3][-25]{%
	\rotatebox{#1}{\makebox[0pt]{\makebox[#2]{\xleaders\hbox{$\cdot$\hskip#3}\hfill\kern0pt}}}%
}
\newcommand{\overbar}[1]{\mkern 4mu\overline{\mkern-4mu#1\mkern-4mu}\mkern 4mu}

\newtheorem{lemma}{Lemma}

\newtheorem{theorem}{Theorem}
\newtheorem{corollary}{Corollary}
\newtheorem{definition}{Definition}

\usetikzlibrary{arrows,shapes,positioning,calc,decorations.pathreplacing,calligraphy}
\tikzstyle{block} = [draw, rectangle,minimum height=3em, minimum width=3em]
\tikzstyle{sum} = [draw, circle]
\tikzstyle{box}=[rectangle, fill=gray!20, draw, minimum width=1.2cm, minimum height=0.5cm, align=center]

\allowdisplaybreaks
\title{\LARGE \bf
On $\mathbf{\ell_2}$-performance of weakly-hard real-time control systems
}

\author{Marc Seidel, Simon Lang, Frank Allgöwer%
\thanks{F. Allg\"ower thanks the German Research Foundation (DFG) for support of this work within grant AL 316/13-2 and within the German Excellence Strategy under grant EXC-2075 - 285825138; 390740016.}%
\thanks{The authors are with the University of Stuttgart, Institute for Systems Theory and Automatic Control, Stuttgart, Germany, {\tt \small\{seidel,lang,allgower\}@ist.uni-stuttgart.de}}%
}

\begin{document}

\maketitle
\thispagestyle{empty}
\pagestyle{empty}

\begin{abstract}
	This paper considers control systems with failures in the feedback channel, that occasionally lead to loss of the control input signal.
	A useful approach for modeling such failures is to consider window-based constraints on possible loss sequences, for example that at least $r$ control attempts in every window of $s$ are successful.
	A powerful framework to model such constraints are weakly-hard real-time constraints.
	Various approaches for stability analysis and the synthesis of stabilizing controllers for such systems have been presented in the past.
	However, existing results are mostly limited to asymptotic stability and rarely consider performance measures such as the resulting $\ell_2$-gain.
	To address this problem, we adapt a switched system description where the switching sequence is constrained by a graph that captures the loss information.
	We present an approach for $\ell_2$-performance analysis involving linear matrix inequalities (LMI).
	Further, leveraging a system lifting method, we propose an LMI-based approach for synthesizing state-feedback controllers with guaranteed $\ell_2$-performance.
	The results are illustrated by a numerical example.
\end{abstract}

\begin{keywords}
	control over communication, switched systems %
\end{keywords}

\input{sections/intro}
\input{sections/setup}

\input{sections/analysis}

\input{sections/lifted_sys}

\input{sections/synthesis}
\input{sections/example}
\input{sections/conclusion}

\bibliographystyle{ieeetr}
\bibliography{literature}

\end{document}

%% file: sections/intro.tex
\section{INTRODUCTION}
In control systems with an unreliable feedback channel, failures can lead to the occasional loss of the control input signal.
Such losses appear for example in Networked Control Systems (NCS), where the feedback loop is closed via a communication channel that is subject to packet loss \cite{Hespanha2007}.
Similarly, in real-time control applications deadline misses can occur, where the control signal computation does not finish in time and thus no new actuator command is issued \cite{Maggio2020}, \ie the control signal can be considered as lost.
In both scenarios, the system runs in open-loop from time to time whenever the attempt to control the system fails, \eg when a control input sent over the network is lost or its computation is not completed before the deadline.
A number of approaches model these losses as random variables, see \cite{Schenato2007} and references therein.
For such models, only stochastic guarantees can be given, \eg in the mean-square sense.
Further approaches deal with deterministic bounded loss \cite{Xiong2007}.
In this case, classical guarantees in the form of stability in the sense of Lyapunov can be given.
However, these approaches neglect information on the past losses, which is rather conservative.

To avoid this conservatism, a new window-like deterministic packet loss model was suggested in the literature \cite{Bernat2001a}.
As a generalization of bounded loss and scheduling constraints like $(m,k)$-firm deadlines \cite{Hamdaoui1995,Horssen2016}, where at least $m$ out of $k$ consecutive tasks are guaranteed to meet their deadline, the notion of weakly-hard real-time (WHRT) constraints has been proposed \cite{Bernat2001a}.
Originally presented as scheduling constraints for real-time applications, these type of constraints have been used recently to model the packet loss in NCS \cite{Blind2015,Linsenmayer2017,Linsenmayer2021a}.
WHRT constraints pose a window-like constraint, where in a moving time window of certain length at least a specified minimum number of control attempts are successful.
This allows to describe losses more accurately compared to a stochastic description, that allows the losses to be distributed unevenly.
For example, a 10\% loss chance allows that $100$ unsuccessful control attempts are followed by $900$ successful ones, or $10$ unsuccessful ones followed by $90$ successful control attempts, which is vastly different from a control perspective \cite{Bernat2001a,Blind2015}.
The notion of WHRT control systems includes NCS with WHRT packet loss description as well as real-time control systems in which the deadline misses are modeled by a WHRT constraint.

For linear WHRT control systems in the context of NCS, \cite{Blind2015,Linsenmayer2017} present linear matrix inequality (LMI) conditions for asymptotic stability analysis and asymptotic state-feedback stabilization.
The work \cite{Linsenmayer2021a} extends this to an event-triggered setup, switching controllers and output-feedback, while \cite{Hertneck2020,Hertneck2021} deal with asymptotic stability for a nonlinear WHRT control system based on the concept of non-monotonic Lyapunov functions.
Another approach in the area of real-time control systems focuses on asymptotic stability analysis using the constrained joint spectral radius \cite{Maggio2020,Vreman2022b}, which characterizes the maximum asymptotic growth rate over all possible switching sequences.
However, the aforementioned works focus solely on asymptotic stability and lack performance considerations, despite the fact that performance guarantees are crucial from an application point of view.
First progress towards performance analysis using a quadratic state cost function has been made in \cite{Pazzaglia2018,Vreman2021a}, however no theoretical guarantees are given.
In \cite{Horssen2016}, performance guarantees in an LQR setting using LMI conditions are studied.
To the best of our knowledge, apart from quadratic costs in \cite{Horssen2016}, classical control performance in the sense of, \eg $\ell_2$-gains has not been considered so far.

This paper aims at addressing the analysis of the $\ell_2$-performance and the synthesis of state-feedback controllers with guaranteed $\ell_2$-performance for WHRT control systems.
To formally describe them, we adapt the switched system description from \cite{Horssen2016,Blind2015,Linsenmayer2017,Linsenmayer2021a}, where the switching is described by a graph, that captures the WHRT constraint, called the WHRT graph \cite{Linsenmayer2021a}.
A similar idea has been presented in \cite{Schendel2010}, where the authors use an automaton model to keep track of dropouts in the feedback loop.
Leveraging the switched system performance results from \cite{Daafouz2002a} and prior work \cite{Daafouz2002}, we derive sufficient LMI conditions for $\ell_2$-performance analysis of WHRT control systems.
This novel result can also be used for $\ell_2$-performance analysis of general \emph{graph-constrained} switched systems.
In contrast to previous works, this has not been done yet for an $\ell_2$-performance measure, but only for stability.
Moreover, we utilize a classical system lifting method \cite{Chen1995} to derive an LMI-based approach for synthesis of switched and non-switched state-feedback controllers with guaranteed $\ell_2$-performance, which additionally allows for a more efficient performance analysis.
The lifting method is based on combining multiple past inputs and outputs and ``discretizing'' the system at time instants of successful control attempts, which directly enables controller synthesis.

The work at hand is based on \cite{Daafouz2002a, Daafouz2002} and motivated by networked and real-time control setups.
It is also possible to view this from a graph-constrained switched systems perspective in a more general dissipativity-based framework.
We present this approach in our work \cite{Lang2024}.
The results obtained in \cite{Lang2024} can be used for WHRT control systems as well, arriving at conditions for robust quadratic performance which contain the results for nominal $\ell_2$-performance, presented in this paper, as special case.

The remainder of the paper is organized as follows.
In Section~\ref{ch:setup}, we define our setup and state preliminaries.
We then present some first $\ell_2$-performance analysis results in Section~\ref{ch:analysis}.
Section~\ref{ch:liftedSystem} defines the lifting method and gives an improved analysis results for the lifted system.
In Section~\ref{ch:synthesis}, controller synthesis is presented and in Section~\ref{ch:example} we emphasize our findings by a numerical example.
Section~\ref{ch:conclusion} concludes the paper.

\subsection{Notation}
The set of real and natural numbers are denoted $\mathbb{R}$ and $\mathbb{N}$, respectively, and the notation $A \succ 0$ means that $A = A^\top$ is positive definite.
Matrix blocks which can be inferred from symmetry are abbreviated with $\ast$.
A (block) diagonal matrix with the blocks $\lambda_1,...,\lambda_n$ is denoted $\mathrm{diag}(\lambda_1,...,\lambda_n)$.

%% file: sections/setup.tex
\section{SETUP} \label{ch:setup}
In this paper, we consider the linear discrete-time plant
\begin{align}
	\begin{split}
		x_{k+1} &= Ax_k + Bu_k^\mathrm{a} + B^w w_k\\
		z_k &= C x_k + Du_k^\mathrm{a} + D^w w_k
	\end{split} \label{eq:system}
\end{align}
with initial state $x_0 \in \mathbb{R}^{n}$, actuator input $u_k^\mathrm{a} \in \mathbb{R}^{m}$, performance input $w_k \in \mathbb{R}^{q}$, performance output $z_k \in \mathbb{R}^{p}$, and real system matrices.
At every time instant $k \in \mathbb{N}_0 \coloneqq \mathbb{N} \cup \{0\}$, the control input $u_k^c \in \mathbb{R}^{m}$ is computed by means of a linear state-feedback controller as
\begin{align}
	u_k^\mathrm{c} = K x_k, \label{eq:controlInput}
\end{align}
which may be lost, \eg due to transmission failure or a deadline miss.
These losses are described by the binary loss sequence $\mu \coloneqq (\mu_k)_{k \in \mathbb{N}_0}$, where $\mu_k = 1$ indicates a successful control attempt and $\mu_k = 0$ a loss.
Lost control inputs are discarded and not sent or computed again.
Without loss of generality, we assume for simplicity that the first control attempt is always successful, \ie $\mu_0 = 1$.
The actuator input $u_k^\mathrm{a}$ depends on the control input $u_k^\mathrm{c}$ and the loss sequence, but also on the chosen strategy in case of a loss.
In literature, typically the zero strategy (input is set to zero if control attempt is not successful) and the hold strategy (last input is held until a new control input is received) are considered, and none of them is seen to be in general superior to the other \cite{Schenato2009}.
In particular, the two strategies result in the actuator inputs
\begin{equation}
	\begin{alignedat}{2}
			u_k^\mathrm{a} &= \mu_k u_k^\mathrm{c} && \quad \text{for zero strategy} \\
			u_k^\mathrm{a} &= \mu_k u_k^\mathrm{c} + (1-\mu_k) u^a_{k-1} && \quad \text{for hold strategy,}
		\label{eq:actuatorInputZeroAndHold}
	\end{alignedat}
\end{equation}
with some initial condition $u_{-1}^\mathrm{a}$ for the hold strategy.
An overview of the system setup is given in Fig.~\ref{fig:systemSetup}.
\begin{figure}
	\centering
	\input{fig/setup.tex}
	\caption{The WHRT control system with losses in the controller-actuator connection.}
	\label{fig:systemSetup}
\end{figure}

\subsection{WHRT constraints}
We assume that the loss sequence $\mu$ satisfies a WHRT constraint $\lambda$.
These type of constraints can be interpreted as a window specification: Within a moving time window, a minimum number of control attempts has to be successful.
More precisely, the standard type of WHRT constraints is defined as follows.
\begin{definition}[\cite{Bernat2001a}] \label{def:anyWHRTC}
	A loss sequence $\mu$ ``meets any $r$ in $s$ control attempts'' ($r$, $s \in \mathbb{N}$, $r \leq s$), denoted $\mu \vdash \anyWHRTC{r}{s} = \lambda$, if in any window of $s$ consecutive control attempts there are at least $r$ of them successful (in any order).
\end{definition}
An example is provided in Figure~\ref{fig:WHC-example}.
Other types of WHRT constraints focus on the number of successful consecutive control attempts, or on the number of unsuccessful control attempts.
\begin{definition}[\cite{Bernat2001a}] \label{def:rowWHRTC}
	A loss sequence $\mu$ ``meets row $r$ in $s$ control attempts'' ($r$, $s \in \mathbb{N}$, $r \leq s$), denoted $\mu \vdash \rowWHRTC{r}{s} = \lambda$, if in any window of $s$ consecutive control attempts, at least $r$ consecutive of them are successful.
\end{definition}
\begin{definition}[\cite{Bernat2001a}]
	A loss sequence $\mu$ ``misses any $r$ in $s$ control attempts'' ($r$, $s \in \mathbb{N}$, $r \leq s$), denoted $\mu \vdash \overbar{\anyWHRTC{r}{s}} = \lambda$, if in any window of $s$ consecutive control attempts there are not more than $r$ of them unsuccessful (in any order).
\end{definition}
\begin{definition}[\cite{Bernat2001a}]
	A loss sequence $\mu$ ``misses row $r$ in $s$ control attempts'' ($r$, $s \in \mathbb{N}$, $r \leq s$), denoted $\mu \vdash \overbar{\rowWHRTC{r}{s}} = \lambda$, if in any window of $s$ consecutive control attempts, there are not more than $r$ consecutive of them unsuccessful.
\end{definition}
Moreover, some WHRT constraints can be related to each other as weaker or harder \cite{Bernat2001a,Vreman2022}.
\begin{definition}[\cite{Bernat2001a}]
	Given two WHRT constraints $\lambda_1$ and $\lambda_2$, we say that $\lambda_1$ is easier than $\lambda_2$ ($\lambda_2$ is harder than $\lambda_1$), denoted $\lambda_2 \preceq \lambda_1$ if $\mu \vdash \lambda_2 \Rightarrow \mu \vdash \lambda_1$.
\end{definition}
\begin{figure}
	\centering
	\input{fig/WHC-example.tex}
	\caption{An exemplary loss sequence $\mu_k = 1 \, 0 \, 0 \, 1 \, 1 \, 1 \, 0 \hdots$ satisfying the constraint $\lambda = \anyWHRTC{2}{4}$: In every window of length 4, there are at least 2 successful control attempts.}
	\label{fig:WHC-example}
\end{figure}
The combination of \eqref{eq:system}, \eqref{eq:controlInput}, \eqref{eq:actuatorInputZeroAndHold}, and the WHRT loss description forms the overall system model.
More formally, we define:
\begin{definition}[WHRT control system]
	The WHRT control system is defined as the system resulting from \eqref{eq:system}, \eqref{eq:controlInput}, and \eqref{eq:actuatorInputZeroAndHold}, whose control input losses are described by the sequence $\mu$ that satisfies a given WHRT constraint, \ie $\mu \vdash \lambda$.
\end{definition}
WHRT control systems occur frequently in, \eg real-time control applications \cite{Maggio2020,Pazzaglia2018}, but also in NCS \cite{Xiong2007,Blind2015}.
There exist techniques to guarantee the satisfaction of WHRT constraints such setups, see for example \cite{Ahrendts2018}.

\subsection{Switched system}
To cope with the WHRT control system we will rewrite it as a constrained switched (linear) system (CSS), where the loss sequence is constrained by a WHRT constraint.
This type of systems has been used in the past to deal with similar setups, \eg in \cite{Horssen2016,Blind2015,Linsenmayer2017,Linsenmayer2021a}, and proves to be useful in ours as well.

The CSS takes the form
\begin{align}
	\begin{split}
		x_{k+1} &= \mathcal{A}_{\mu_k}^\mathrm{cl} x_k + \mathcal{B}_{\mu_k}^w w_k\\
		z_k &= \mathcal{C}_{\mu_k}^\mathrm{cl} x_k + \mathcal{D}_{\mu_k}^w w_k,
	\end{split} \label{eq:switchedSystem}
\end{align}
where the real system matrices depend on the loss sequence $\mu$.
Thus, the CSS has two modes: $\mu_k = 1$ for a successful control attempt and $\mu_k = 0$ representing a loss at time instant $k$, \eg $\mathcal{A}^{cl}_{\mu_k} \in \{ \mathcal{A}_{0}^\mathrm{cl}, \mathcal{A}_{1}^\mathrm{cl}\}$.
Note that the system matrices can be chosen such that \eqref{eq:switchedSystem} is an interchangeable representation of the WHRT control system.
For example,
\begin{equation} \label{eq:ssMatricesZero}
	\begin{alignedat}{3}
		\mathcal{A}_0^\mathrm{cl} &= A \qquad & \mathcal{A}_1^\mathrm{cl} &= A+BK \qquad & \mathcal{B}^w &= B^w \\
		\mathcal{C}_0^\mathrm{cl} &= C \qquad & \mathcal{C}_1^\mathrm{cl} &= C+DK \qquad & \mathcal{D}^w &= D^w
	\end{alignedat}
\end{equation}
is obtained for the zero strategy, \cf \cite{Blind2015,Linsenmayer2017}.
For the respective hold strategy matrices we refer to \cite{Blind2015,Linsenmayer2017}, which require an augmented state vector with the additional auxiliary state $u_{k-1}$ for storing the last applied input.
We will later state the full system matrices for both strategies in the lifted system formulation.

\subsection{$\ell_2$-performance}
The $\ell_2$-gain is a classical performance measure for control systems, that can be interpreted as the worst-case energy amplification from a performance input (\eg a disturbance) $w_k$ to a performance output $z_k$ \cite{Khalil2002}.
More formally, we define $\ell_2$-performance and the $\ell_2$-gain as follows.
\begin{definition}[$\ell_2$-performance] \label{def:l2}
	The CSS \eqref{eq:switchedSystem} has $\ell_2$-performance with gain $\gamma$ if it is asymptotically stable and for $x_0 = 0$
	\begin{align} \label{eq:defPerformance}
		\sum_{k=0}^{\infty} z_k^\top z_k < \gamma^2 \sum_{k=0}^{\infty} w_k^\top w_k \quad \parbox{7em}{$\forall w \in \ell_2, w \neq 0$,\\ $\forall \mu \vdash \lambda$,}
	\end{align}
	where $\ell_2$ denotes the set of all square-summable signals, \ie all $w$ for which $\sum_{k=0}^{\infty} w_k^\top w_k < \infty$.
	The smallest $\gamma$ such that \eqref{eq:defPerformance} is still satisfied is called the $\ell_2$-gain of the system.
\end{definition}
In this work, we are interested in the $\ell_2$-performance of the WHRT control system.
For that, we will first study the $\ell_2$-performance of the CSS \eqref{eq:switchedSystem} and then transfer the results to the WHRT control system.

%% file: fig/setup.tex
\begin{tikzpicture}[auto,>=latex,thick]
	\node [block] (sys) {Plant \eqref{eq:system}};
	\node [left=of sys](drop) {};
	\node [block, left=of drop] (K) {Controller $K$};
	\draw (K.east) --  node[below, pos=0.5] {$u_k^c$} (drop.west);
	\draw (drop.west) -- node[above, pos=0.5] {$\mu_k$} (drop.north east);
	\draw [-latex] (drop.east) --  node[below, pos=0.5] {$u_k^a$} (sys.west);
	\draw [-latex] (sys.north) +(0,0.7) -- node[right, pos=0.25] {$w_k$} (sys.north);
	\draw [-latex] (sys.east) -- node[above, pos=0.5] {$z_k$} +(0.7,0);
	\draw [-latex] (sys.south) |- node[below, pos=0.75] {$x_k$} ($(K.south west)+(-0.5,-0.5)$) |- (K.west);
\end{tikzpicture}

%% file: fig/WHC-example.tex
\begin{tikzpicture}
	\def\numAttempt{7}
	\def\spacingAxis{1.0}
	\def\sideLengthBraces{0.0} %

	\begin{scope}[shift={(0,0)}]
		\draw[thick, ->] (-0.5*\spacingAxis,0) -- (\numAttempt*\spacingAxis-0.5*\spacingAxis,0);
		\node[right] at (\spacingAxis*\numAttempt-0.5*\spacingAxis,0) {$k$};

		\draw[dotted,gray,thick] ($(0*\spacingAxis,-0.1)+(0,-1.3em)$) -- ($(0*\spacingAxis,-0.1)+(0,-1.9em)$);
		\draw[dotted,gray,thick] ($(1*\spacingAxis,-0.1)+(0,-1.3em)$) -- ($(1*\spacingAxis,-0.1)+(0,-4.3em)$);
		\draw[dotted,gray,thick] ($(2*\spacingAxis,-0.1)+(0,-1.3em)$) -- ($(2*\spacingAxis,-0.1)+(0,-6.5em)$);
		\draw[dotted,gray,thick] ($(3*\spacingAxis,-0.1)+(0,-1.3em)$) -- ($(3*\spacingAxis,-0.1)+(0,-8.8em)$);
		\draw[dotted,gray,thick] ($(4*\spacingAxis,-0.1)+(0,-1.3em)$) -- ($(4*\spacingAxis,-0.1)+(0,-4.3em)$);
		\draw[dotted,gray,thick] ($(5*\spacingAxis,-0.1)+(0,-1.3em)$) -- ($(5*\spacingAxis,-0.1)+(0,-6.5em)$);
		\draw[dotted,gray,thick] ($(6*\spacingAxis,-0.1)+(0,-1.3em)$) -- ($(6*\spacingAxis,-0.1)+(0,-8.8em)$);

		\foreach \k / \label in {0/1, 1/0, 2/0, 3/1, 4/1, 5/1, 6/0} {
			\ifnum \label=0
				\node[below] at (\k*\spacingAxis,-0.1) (attempt\k) {0};
			\else
				\node[below] at (\k*\spacingAxis,-0.1) (attempt\k) {1};
			\fi
			
			\draw[thick] (\k*\spacingAxis,0.1) -- (\k*\spacingAxis,-0.1);
		}
		\node[below] at (-0.7*\spacingAxis,-0.15) {$\mu=$};

		\begin{scope}[shift={(0,-0.3em)}]
			\draw[decorate,thick,decoration={calligraphic brace,amplitude=5pt,mirror}] (0*\spacingAxis-\sideLengthBraces*\spacingAxis,-0.65) -- node[below,yshift=-0.3em] {2 successes} (3*\spacingAxis+\sideLengthBraces*\spacingAxis,-0.65);
		\end{scope}
		\begin{scope}[shift={(0,-2.6em)}]
			\draw[decorate,thick,decoration={calligraphic brace,amplitude=5pt,mirror}] (1*\spacingAxis-\sideLengthBraces*\spacingAxis,-0.65) -- node[below,yshift=-0.3em] {2 successes} (4*\spacingAxis+\sideLengthBraces*\spacingAxis,-0.65);
		\end{scope}
		\begin{scope}[shift={(0,-4.9em)}]
			\draw[decorate,thick,decoration={calligraphic brace,amplitude=5pt,mirror}] (2*\spacingAxis-\sideLengthBraces*\spacingAxis,-0.65) -- node[below,yshift=-0.3em] {3 successes} (5*\spacingAxis+\sideLengthBraces*\spacingAxis,-0.65);
		\end{scope}
		\begin{scope}[shift={(0,-7.2em)}]
			\draw[decorate,thick,decoration={calligraphic brace,amplitude=5pt,mirror}] (3*\spacingAxis-\sideLengthBraces*\spacingAxis,-0.65) -- node[below,yshift=-0.3em] {3 successes} (6*\spacingAxis+\sideLengthBraces*\spacingAxis,-0.65);
		\end{scope}
		
	\end{scope}

\end{tikzpicture}

%% file: sections/analysis.tex
\section{$\ell_2$-PERFORMANCE ANALYSIS FOR CSSs} \label{ch:analysis}
In this section, we derive a sufficient condition to verify that the CSS \eqref{eq:switchedSystem} has $\ell_2$-performance with a certain gain $\gamma$.
There already exist $\ell_2$-performance results for switched systems in the literature \cite{Daafouz2002a,Fang2004}.
However, these results are limited to arbitrary switching between the modes of the switched system.
Since in our setup the switching sequence is not arbitrary but constrained by the WHRT constraint, using arbitrary switching results induces conservatism.
We leverage these results by explicitly taking the constrained switching into consideration.

\subsection{Graph representation}
To formalize the notion of constrained switching, a particular concept has been proven useful in several previous works \cite{Horssen2016,Linsenmayer2017,Linsenmayer2021a,Philippe2016}: representing the WHRT constraint $\lambda$ as an automaton described by a graph, that generates the allowed switching sequences.

\begin{definition}[WHRT graph] \label{def:switchingGraph}
	For a WHRT constraint $\lambda$, the corresponding WHRT graph $\mathcal{G}$ satisfies the following properties:
	\begin{enumerate}[label=(\alph*)]
		\item \label{prop:WHRTgraph-mathDef} It is a labeled directed graph $\mathcal{G} = (\mathcal{V},\mathcal{E})$ with the node set $\mathcal{V} = \{v_1,...,v_{n_\mathcal{V}}\}$ and the edge set $\mathcal{E} = \{e_1,...,e_{n_\mathcal{E}}\}$ with $e_p = (i_p,j_p,l_p)$, where $e_p \in \mathcal{E}$ if there exists an edge with label $l_p$ from node $v_{i_p}$ to node $v_{j_p}$.
		\item \label{prop:WHRTgraph-labels} The labels $l_p$ take values in the set $\{0,1\}$, representing a loss of the control input signal or a successful control attempt.
		\item \label{prop:WHRTgraph-sequenceSatisfaction} All sequences $\mu \vdash \lambda$ can be generated by the graph.
		\item \label{prop:WHRTgraph-outAndInEdges} Every node of $\mathcal{G}$ has at least one incoming and one outgoing edge.
	\end{enumerate}
\end{definition}
As a consequence, we can represent the loss sequence $\mu$ by moving along the edges of the corresponding graph.
The labels thereby characterize the current mode of the CSS, whereas the graph specifies how the modes can be concatenated such that $\mu \vdash \lambda$.
Based on Definition~\ref{def:switchingGraph}, define further the indicator function $\eta_i\colon \mathbb{N}_0 \rightarrow \{0,1\}$, $i = 1,...,n_\mathcal{V}$ with
\begin{align} \label{eq:indicatorFcn-nonLifted}
	\eta_i(k) = \begin{cases}
		1, \quad \text{for $i = i_p$ with $\mu_k = l_p$} \\
		0, \quad \text{otherwise},
	\end{cases}
\end{align}
\ie $\eta$ indicates the initial node of the current edge, that is the current mode $\mu_k$ at time instant $k$.
Note that it is always possible to generate such a graph for any given WHRT constraint.
We will discuss the corresponding details in Section~\ref{ch:liftedSystem}.
The size of the graph is discussed and determined in \cite{Vreman2022}, which can get large for some WHRT constraints.
The lifting method presented later in Section~\ref{ch:liftedSystem} is also able to reduce the graph size.

\subsection{$\ell_2$-performance analysis result}
Given the previous graph definition, we can now state our first result.
We use \cite[Theorem~2]{Daafouz2002a} as a basis and modify it to account for the constrained switching.
\begin{theorem}[$\ell_2$-performance analysis] \label{thm:analysis-nonLifted}
	The CSS \eqref{eq:switchedSystem}, whose switching is captured by a graph $\mathcal{G}$ according to Definition~\ref{def:switchingGraph}, has $\ell_2$-performance with gain $\gamma$ if there exist symmetric matrices $S_i \in \mathbb{R}^{n \times n}$ and matrices $G_i \in \mathbb{R}^{n \times n}$, $i = 1,...,n_\mathcal{V}$, such that
	\begin{align}
		\begin{bmatrix}
			G_i + G_i^\top - S_i & \ast & \ast & \ast \\
			0 & \gamma I & \ast & \ast \\
			\mathcal{A}_l^\mathrm{cl} G_i & \mathcal{B}_l^w & S_j & \ast \\
			\mathcal{C}_l^\mathrm{cl} G_i & \mathcal{D}_l^w & 0 & \gamma I
		\end{bmatrix} \succ 0 \quad \forall (i,j,l) \in \mathcal{E}. \label{eq:lmiAnalysis-nonLifted} \raisetag{1\baselineskip}
	\end{align}
	A Lyapunov function is then given by
	\begin{align} \label{eq:lyapFcn-nonLifted}
		V(x_k) = x_k^\top \left( \sum_{i=1}^{n_\mathcal{V}} \eta_i(k) S_i^{-1} \right)x_k.
	\end{align} %
\end{theorem}
\begin{proof}
	Since every node has at least one incoming and one outgoing edge, \eqref{eq:lmiAnalysis-nonLifted} guarantees $S_i^{-1} \succ 0 ~ \forall i$.
	Note that \eqref{eq:lmiAnalysis-nonLifted} implies
	\begin{align}
		\begin{bmatrix}
			G_i + G_i^\top - S_i & \ast \\
			\mathcal{A}_l^\mathrm{cl} G_i & S_j & \\
		\end{bmatrix} \succ 0 \quad \forall (i,j,l) \in \mathcal{E},
	\end{align}
	which implies asymptotic stability by \cite[Corollary~9]{Linsenmayer2017}.
	The remaining proof is adapted from \cite[Theorem~2]{Daafouz2002a}, but modified to cope with the graph induced constrained switching.
	Applying the congruence transformation $\mathrm{diag}(0,\gamma^{-1}I,0,\gamma^{-1}I)$ on the LMI \eqref{eq:lmiAnalysis-nonLifted}, it is equivalent to
	\begin{align*}
		\begin{bmatrix}
			\bar{G}_i + \bar{G}_i^\top - \bar{S}_i & \ast \\
			\bar{A}_l^\mathrm{cl} \bar{G}_i & \bar{S}_j
		\end{bmatrix} \succ 0 \quad \forall (i,j,l) \in \mathcal{E},
	\end{align*}
	where
	\begin{alignat*}{3}
		\bar{G}_i &\coloneqq \begin{bmatrix}
			G_i & 0 \\
			0 & \gamma^{-1} I
		\end{bmatrix} \quad && \bar{S}_i &&\coloneqq \begin{bmatrix}
			S_i & 0 \\
			0 & \gamma^{-1} I
		\end{bmatrix} \\
		\bar{\mathcal{A}}_l^\mathrm{cl} &\coloneqq \begin{bmatrix}
			\mathcal{A}_l^\mathrm{cl} & \mathcal{B}_l^w\\
			\gamma^{-1}\mathcal{C}_l^\mathrm{cl} & \gamma^{-1} \mathcal{D}_l^w
		\end{bmatrix} \quad && \bar{S}_j &&\coloneqq \begin{bmatrix}
			S_j & 0 \\
			0 & \gamma^{-1} I
		\end{bmatrix}.
	\end{alignat*}
	Following the same steps as in the proof of \cite[Theorem~2]{Daafouz2002} (step iii) $\Rightarrow$ ii)) yields
	\begin{align}
		\begin{bmatrix} \bar{S}_i^{-1} & \ast \\ \bar{S}_j^{-1} \bar{A}_l^\mathrm{cl} & \bar{S}_j^{-1} \end{bmatrix} \succ 0 \quad \forall (i,j,l) \in \mathcal{E}.
	\end{align}
	Using the Schur complement and resubstituting the $\bar{(\cdot)}$ variables, we obtain for all $(i,j,l) \in \mathcal{E}$
	\begin{align*}
		\begin{bmatrix} \ast \end{bmatrix}^\top
		\begin{bmatrix} S_j^{-1} & 0 \\ 0 & \gamma I\end{bmatrix}
		\begin{bmatrix} \mathcal{A}_l^\mathrm{cl} & \mathcal{B}_l^w \\ \gamma^{-1}\mathcal{C}_l^\mathrm{cl} & \gamma^{-1} \mathcal{D}_l^w\end{bmatrix}
		- \begin{bmatrix} S_i^{-1} & 0 \\ 0 & \gamma I \end{bmatrix} \prec 0.
	\end{align*}
	Since the graph represents the constrained switching with the labels taking the values in $\{0,1\}$, we can set $l_p = \mu_k$, and utilize \eqref{eq:lyapFcn-nonLifted}, \eqref{eq:indicatorFcn-nonLifted} for the initial node $i_p$ of the edge at time instant $k$ and for the receiving node $j_p$ of the edge at time instant $k+1$.
	This results in
	\begin{align*}
		\begin{bmatrix} \ast \end{bmatrix}^\top
		\hspace{-0.3em} \begin{bmatrix} P_{i,k+1} & 0 \\ 0 & \gamma I \end{bmatrix}
		\hspace{-0.3em}\begin{bmatrix} \mathcal{A}_{\mu_k}^\mathrm{cl} & \mathcal{B}_{\mu_k}^w \\ \gamma^{-1} \mathcal{C}_{\mu_k}^\mathrm{cl} & \gamma^{-1} \mathcal{D}^w_{\mu_k} \end{bmatrix}
		\hspace{-0.2em} - \hspace{-0.2em} \begin{bmatrix} P_{i,k} & 0 \\ 0 & \gamma I\end{bmatrix} \prec 0.
	\end{align*}
	We multiply this equation with $\begin{bmatrix} x_k^\top & w_k^\top \end{bmatrix}$ from the left and its transpose from the right and use \eqref{eq:lyapFcn-nonLifted} and \eqref{eq:switchedSystem} to end up with $V(x_{k+1}) - V(x_k) < \gamma w_k^\top w_k - \gamma^{-1} z_k^\top z_k$ $\forall w_k \in \ell_2$, $\forall x_k$, which is a sufficient condition for \eqref{eq:switchedSystem} having $\ell_2$-performance with gain $\gamma$, see, \eg \cite{Daafouz2002a,Fang2004}.
\end{proof}
We remark that \eqref{eq:lyapFcn-nonLifted} defines a switched Lyapunov function.
This is a commonly used concept \cite{Daafouz2002a,Daafouz2002,Linsenmayer2017}, that improves the feasibility of the analysis problem, as not one common Lyapunov function is required.
Instead, the different $S_i^{-1}$ define scenario-specific Lyapunov functions, that can be different depending on the node of the WHRT graph that is induced by the dropout sequence.
The same concept is used in \cite{Schendel2010}, where the authors design dropout-dependent Lyapunov functions.

Moreover note that Theorem~\ref{thm:analysis-nonLifted} can be used to analyze the $\ell_2$-performance of an arbitrary CSS, whose switching sequence can be captured by a graph.
Previous results only captured the $\ell_2$-performance of \emph{unconstrained} switched systems.
In our setup, the WHRT control system can be formulated as a CSS \eqref{eq:switchedSystem}, see \eqref{eq:ssMatricesZero}, and thus Theorem~\ref{thm:analysis-nonLifted} can directly be used to analyze a given WHRT control system not only regarding its stability as in \cite{Linsenmayer2017}, but additionally regarding its $\ell_2$-performance.
Moreover, we can minimize $\gamma$ under the LMI constraints \eqref{eq:lmiAnalysis-nonLifted} to obtain an upper bound on the $\ell_2$-gain of the CSS.
However, the underlying graph and thus the number of decision variables of the LMIs \eqref{eq:lmiAnalysis-nonLifted} may be relatively large, since we need to solve one LMI per edge, and have $n^2$ decision variables in $G_i$ per node plus $\frac{1}{2}(n^2+n)$ decision variables in $S_i = S_i^\top$ per node.
Therefore, the conditions of Theorem~\ref{thm:analysis-nonLifted} might be computationally expensive to solve.
Moreover, controller synthesis becomes difficult, because it is not possible to isolate $K$ in \eqref{eq:lmiAnalysis-nonLifted}.
The reason is that the matrices $\mathcal{A}_{\mu_k}^\mathrm{cl}$ and $\mathcal{C}_{\mu_k}^\mathrm{cl}$ might not be linear in $K$, dependent on the mode.
For example in \eqref{eq:ssMatricesZero}, $\mathcal{A}_{0}^\mathrm{cl}$ is independent of $K$ since the computed control signal does not arrive at the actuator whenever a loss occurs and thus the system runs open-loop.
In contrast, $K$ appears linearly in $\mathcal{A}_{1}^\mathrm{cl}$, \ie when the control attempt is successful.
In the next section, we show how we can address these issues by lifting the system in an appropriate way.

%% file: sections/lifted_sys.tex
\section{LIFTED SYSTEM} \label{ch:liftedSystem}
In this section, we apply a lifting method to the CSS \eqref{eq:switchedSystem} to enable controller synthesis.
Furthermore, by reducing the graph size, the computational complexity of our approach can be decreased, leading to a more efficient analysis.

\subsection{Main idea and lifting}
The fundamental concept is to ``discretize'' the system at time instants of successful control attempts, similarly as in \cite{Linsenmayer2017}.
This approach was called \emph{alternative discretization} and constitutes a special case of the lifting presented in this section, which is adapted from \cite{Chen1995}.
To formalize, define $\tau \coloneqq (\tau_{\tilde{k}})_{\tilde{k} \in \mathbb{N}_0}$ as in \cite{Linsenmayer2017} as the sequence of time indices of successful control attempt instants $k$, and $\alpha \coloneqq (\alpha_{\tilde{k}})_{\tilde{k} \in \mathbb{N}_0}$ as the sequence counting the number of losses between two successful control attempts.
The sequence $\alpha$ plays an important role in the lifting and its elements take values in the finite set $\{0, ..., s-1\}$, since we assumed $s$ to be bounded and hence there can only be a finite amount of consecutive losses.
As an example, for $\mu = (1\,0\,1\,1\,0\,0\,1 ...)$ the corresponding sequences for the lifting are $\tau = (0\,2\,3\,6 ...)$ and $\alpha = (1\,0\,2 ...)$.

The goal is to find an alternative representation of \eqref{eq:switchedSystem}, but based on the successful control attempt time instants.
It is therefore defined at the time instants $\tau_{\tilde{k}}$ only.
All variables which refer to the lifted system are marked with $\tilde{(\boldsymbol{\cdot})}$, \eg $\tilde{x}$ represents the lifted state of $x$.
In \cite{Linsenmayer2017}, this alternative representation is relatively straightforward, because autonomous systems with no performance in- and output are considered therein.
In contrast, our setup features such an in- and output, which cannot be discarded at time instants between successful control attempts without changing the $\ell_2$-performance of the system.
For this reason, we lift the input and output as follows
\begin{align}
	\begin{split}
		\tilde{w}_{\tilde{k}} &= \begin{bmatrix}
			w^\top_{\tau_{{\tilde{k}}-1}} & \cdots & w^\top_{\tau_{\tilde{k}}-1}
		\end{bmatrix}^\top\\
		\tilde{z}_{\tilde{k}} &= \begin{bmatrix}
			z^\top_{\tau_{{\tilde{k}}-1}} & \cdots & z^\top_{\tau_{\tilde{k}}-1}
		\end{bmatrix}^\top.
	\end{split}
\end{align}
Note that at each time instant $\tau_{\tilde{k}}$, the lifted in- and output $\tilde{w}_{\tilde{k}}$ and $\tilde{z}_{\tilde{k}}$ consist of past in- and outputs.
Since we are interested in an equivalent formulation of \eqref{eq:switchedSystem}, the lifted and the non-lifted system states are required to coincide at these time instants, \ie $x_k = \tilde{x}_{\tau_{\tilde{k}}}$.
The lifted CSS is then defined as
\begin{align}
	\begin{split} \label{eq:switchedSystem-lifted}
		\tilde{x}_{{\tilde{k}}+1} &= \tilde{\mathcal{A}}_{\alpha_{\tilde{k}}}^\mathrm{cl} \tilde{x}_{\tilde{k}} + \tilde{\mathcal{B}}_{\alpha_k}^w \tilde{w}_{\tilde{k}}\\
		\tilde{z}_{\tilde{k}} &= \tilde{\mathcal{C}}_{\alpha_{\tilde{k}}}^\mathrm{cl} \tilde{x}_{\tilde{k}} + \tilde{\mathcal{D}}_{\alpha_{\tilde{k}}}^w \tilde{w}_{\tilde{k}}
	\end{split}
\end{align}
with the closed-loop matrices %
\begin{align} \label{eq:switchedSystemClMatrices-lifted}
	\begin{split}
		\tilde{\mathcal{A}}_{\alpha_{\tilde{k}}}^\mathrm{cl} &\coloneqq \tilde{\mathcal{A}}_{\alpha_{\tilde{k}}} + \tilde{\mathcal{B}}_{\alpha_{\tilde{k}}} K \\
		\tilde{\mathcal{C}}_{\alpha_{\tilde{k}}}^\mathrm{cl} &\coloneqq \tilde{\mathcal{C}}_{\alpha_{\tilde{k}}} + \tilde{\mathcal{D}}_{\alpha_{\tilde{k}}} K.
	\end{split}
\end{align}
Since the number of time instants between two successful control attempts is not constant, the lifted CSS has varying in- and output dimensions, depending on the current mode $\alpha_{\tilde{k}}$.
For WHRT control systems, we state the lifted system matrices explicitly in the next subsection.

\subsection{Application to WHRT control systems}
The modes of the lifted system \eqref{eq:switchedSystem-lifted} are not dependent on the loss sequence $\mu$ anymore, because it is defined at time instants of successful control attempts only.
Instead, the switching is now based on the sequence $\alpha$, whose elements take values in $\{0, ...,  s-1\}$.
Therefore, we need to redefine the WHRT graph, similar as in \cite{Linsenmayer2021a}.
\begin{definition}[Lifted WHRT graph] \label{def:WHRTgraph}
	For a WHRT constraint $\lambda$, the lifted WHRT graph $\tilde{\mathcal{G}}$ is defined as in Definition~\ref{def:switchingGraph}, but redefining
	\begin{enumerate}
		\item[$\tilde{\text{\ref{prop:WHRTgraph-labels}}}$] The labels $l_p$ take values in the set $\{0, ..., s-1\}$, representing the number of losses between successful control attempts.
	\end{enumerate}
\end{definition}
The lifted WHRT graph has $\tilde{n}_\mathcal{V}$ nodes and $\tilde{n}_\mathcal{E}$ edges.
Further, the indicator function \eqref{eq:indicatorFcn-nonLifted} is redefined to $\tilde{\eta}_i(\tilde{k})$ by replacing $\mu_k$ by $\alpha_{\tilde{k}}$ and $n_\mathcal{V}$ by $\tilde{n}_\mathcal{V}$.
Similar to the non-lifted graph, the $\alpha$-sequence can be generated by moving along the edges of $\tilde{\mathcal{G}}$.
It is always possible to generate a lifted WHRT graph for any given WHRT constraint.
An algorithm for its automatic generation has been presented in \cite{Linsenmayer2021a} and an implementation is referenced therein, while an algorithm for generating a non-lifted WHRT graph is given in \cite{Vreman2022}.
Therein, the authors furthermore present a method for combining different WHRT constraints into a joint graph.
Further, a (non-lifted) WHRT graph can be generated from the lifted version as follows.
Any edge with $l_p \geq 1$ represents $l_p$ consecutive losses followed by one successful control attempt.
Replacing these edges by a series of $l_p$ edges with label $0$ followed by one with label $1$, including the necessary nodes in-between, leads to a (non-lifted) WHRT graph for a given WHRT constraint.
An example WHRT graph and its corresponding lifted version can be found in Fig.~\ref{fig:example-WHRTgraph}, where the relation between both graphs can be observed.
\begin{figure}
	\centering
	\input{fig/exampleWHRTgraph.tex}
	\caption{Left: non-lifted WHRT graph with labels representing the loss sequence $\mu$; right: its lifted counterpart with labels representing $\alpha$.}
	\label{fig:example-WHRTgraph}
\end{figure}

We are now able to state the WHRT control system as a lifted CSS of the form \eqref{eq:switchedSystem-lifted} with \eqref{eq:switchedSystemClMatrices-lifted}.
Performing the calculations for the WHRT control system (equations \eqref{eq:system} \textendash~\eqref{eq:actuatorInputZeroAndHold}), ``discretized'' at time instants of successful control attempts, one obtains using the sequence $\alpha$
\begin{align}
	\begin{split} \label{eq:matricesLiftedSysZero} \raisetag{3.0\baselineskip}
		\tilde{\mathcal{B}}_{\alpha_{\tilde{k}}} &= A^{\alpha_{\tilde{k}}} B \\
		\tilde{\mathcal{D}}_{\alpha_{\tilde{k}}} &= \begin{cases}
			\text{for $\alpha_{\tilde{k}} = 0$:} ~ D \\
			\text{for $\alpha_{\tilde{k}} \geq 1$:}\\
			\, \begin{bmatrix} D^\top & (C A^0 B)^\top & \cdots & (CA^{\alpha_{\tilde{k}}-1}B)^\top \end{bmatrix}^\top
		\end{cases}
	\end{split}
\end{align}
for the zero strategy and
\begin{align}
	\begin{split} \label{eq:matricesLiftedSysHold} \raisetag{3.5\baselineskip}
		\tilde{\mathcal{B}}_{\alpha_{\tilde{k}}} &= \sum_{i=0}^{\alpha_{\tilde{k}}}A^i B\\
		\tilde{\mathcal{D}}_{\alpha_{\tilde{k}}} &= \begin{cases}
			\text{for $\alpha_{\tilde{k}} = 0$:} ~ D \\
			\text{for $\alpha_{\tilde{k}} \geq 1$:}\\
			\, \begin{bmatrix} D^\top & (CB+D)^\top & \hspace{-0.2em} \cdots & (C \hspace{-0.5em} \sum\limits_{i=0}^{\alpha_{\tilde{k}}-1} \hspace{-0.5em} A^i B + D)^\top  \end{bmatrix}^\top
		\end{cases}
	\end{split}
\end{align}
for the hold strategy, while
\begin{align}
	\begin{split} \label{eq:matricesLiftedSysBoth} \raisetag{5.5\baselineskip}
		\tilde{\mathcal{A}}_{\alpha_{\tilde{k}}} &= A^{\alpha_{\tilde{k}}+1} \\
		\tilde{\mathcal{C}}_{\alpha_{\tilde{k}}} &= \begin{bmatrix} (C A^0)^\top & (CA)^\top & \cdots & (CA^{\alpha_{\tilde{k}}})^\top \end{bmatrix}^\top\\
		\tilde{\mathcal{B}}^w_{\alpha_{\tilde{k}}} &= \begin{bmatrix} A^{\alpha_{\tilde{k}}} B^w & \cdots & A B^w & A^0 B^w\end{bmatrix} \\
		\tilde{\mathcal{D}}^w_{\alpha_{\tilde{k}}} &= \begin{cases}
			\text{for $\alpha_{\tilde{k}} = 0$:} \quad D^w \\
			\text{for $\alpha_{\tilde{k}} \geq 1$:}
			~\begin{bmatrix}
				D^w & 0 & \cdots & 0 \\
				C B^w & \multicolumn{2}{c}{\smash{\raisebox{-0.8\normalbaselineskip}{\diagdots[-30]{6.70em}{.5em}}}} & \vdots \\
				\vdots & \multicolumn{1}{c}{\smash{\raisebox{0.2\normalbaselineskip}{\diagdots[-30]{4.0em}{.5em}}}} & \multicolumn{1}{c}{\smash{\raisebox{1.7\normalbaselineskip}{\diagdots[-30]{4.0em}{.5em}}}} & 0 \\
				C A^{\alpha_{\tilde{k}}-1} B^w & \mathclap{\cdots} & CB^w & D^w
			\end{bmatrix}
		\end{cases}
	\end{split}
\end{align}
are identical for both strategies.
These matrices also appear for lifting general (non-switched) discrete-time systems, \cf \cite[Chapter~8.2]{Chen1995}.
Note that compared to \cite{Linsenmayer2021a} and previous works, due to our lifting no auxiliary state $u_{k-1}^\mathrm{a}$ is required for the hold strategy, thus reducing the state-space dimension.

Note that we can handle one- or multi-step delays in the feedback channel \cite{Maggio2020} with our framework.
For that, augment the system state $x_k$ with the delayed inputs $[u_{k-1}^\mathrm{a}, u_{k-2}^\mathrm{a}, ...]$ and adapt the system matrices in \eqref{eq:system} accordingly.
The later synthesized controller will additionally to $x_k$ depend on the delayed inputs, which have to be known to the controller, \eg by acknowledgment mechanisms of the communication channel.

\subsection{$\ell_2$-performance for the lifted system}
Observe that by definition of $\tilde{w}$ and $\tilde{z}$
\begin{align*}
	\sum_{k=0}^{\infty} z_k^\top z_k = \sum_{\tilde{k}=0}^{\infty} \tilde{z}_{\tilde{k}}^\top \tilde{z}_{\tilde{k}} \quad \text{and} \quad \sum_{k=0}^{\infty} w_k^\top w_k = \sum_{\tilde{k}=0}^{\infty} \tilde{w}_{\tilde{k}}^\top \tilde{w}_{\tilde{k}}.
\end{align*}
Therefore, lifting the system loses no input/output information, the $\ell_2$-gain stays the same, also because asymptotic stability of the lifted system \eqref{eq:switchedSystem-lifted} carries over to the non-lifted one \eqref{eq:switchedSystem} \cite{Linsenmayer2017}.
Further, $\ell_2$-performance guarantees given for \eqref{eq:switchedSystem-lifted} carry over to \eqref{eq:switchedSystem}.
The following Lemma states the equivalence.
\begin{lemma} \label{thm:l2gainEquivalence}
	The CSS \eqref{eq:switchedSystem} and its lifted counterpart \eqref{eq:switchedSystem-lifted} have the same $\ell_2$-performance gain $\gamma$.
\end{lemma}
We can now state a similar $\ell_2$-performance analysis result for the lifted system \eqref{eq:switchedSystem-lifted} as for the non-lifted system \eqref{eq:switchedSystem}.
\begin{theorem}[Lifted system $\ell_2$-performance analysis] \label{thm:analysis-lifted}
	The lifted CSS \eqref{eq:switchedSystem-lifted}, whose switching is captured by a graph $\tilde{\mathcal{G}}$ according to Definition~\ref{def:WHRTgraph}, has $\ell_2$-performance with gain $\gamma$ if there exist symmetric matrices $S_i \in \mathbb{R}^{n \times n}$ and matrices $G_i \in \mathbb{R}^{n \times n}$, $i = 1,...,\tilde{n}_\mathcal{V}$, such that
	\begin{align} \label{eq:lmiAnalysis-lifted} \raisetag{1\baselineskip}
		\begin{bmatrix}
			G_i + G_i^\top - S_i & \ast & \ast & \ast \\
			0 & \gamma I & \ast & \ast \\
			\tilde{\mathcal{A}}_l^\mathrm{cl} G_i & \tilde{\mathcal{B}}_l^w & S_j & \ast \\
			\tilde{\mathcal{C}}_l^\mathrm{cl} G_i & \tilde{\mathcal{D}}_l^w & 0 & \gamma I
		\end{bmatrix} \succ 0 \quad \forall (i,j,l) \in \mathcal{E}.
	\end{align}
	A Lyapunov function is then given by
	\begin{align} \label{eq:lyapFcn-lifted}
		V(\tilde{x}_{\tilde{k}}) = \tilde{x}_{\tilde{k}}^\top \left( \sum_{i=1}^{\tilde{n}_\mathcal{V}} \tilde{\eta}_i(\tilde{k}) S_i^{-1} \right) \tilde{x}_{\tilde{k}}. %
	\end{align}
\end{theorem}
\begin{proof}
	The proof follows the same steps as in Theorem~\ref{thm:analysis-nonLifted}. The dimensions of $\mathcal{B}_{\mu_k}^w$, $\mathcal{C}_{\mu_k}^\mathrm{cl}$, and $\mathcal{D}_{\mu_k}^w$ in the proof of Theorem~\ref{thm:analysis-nonLifted} have to be adapted accordingly.
\end{proof}
\begin{corollary} \label{thm:analysis-WHRTsystem}
	Under the assumptions of Theorem~\ref{thm:analysis-lifted}, the WHRT control system has $\ell_2$-performance with gain $\gamma$.
\end{corollary}
\begin{proof}
	From Theorem~\ref{thm:analysis-lifted} and Lemma~\ref{thm:l2gainEquivalence} we obtain that \eqref{eq:switchedSystem-lifted} and \eqref{eq:switchedSystem} have $\ell_2$-performance with gain $\gamma$. The fact that the WHRT control system can be represented by \eqref{eq:switchedSystem-lifted} completes the proof.
\end{proof}
Due to the lifting, the system matrices $\tilde{\mathcal{B}}_{\alpha_{\tilde{k}}}^w$, $\tilde{\mathcal{C}}_{\alpha_{\tilde{k}}}^\mathrm{cl}$, and $\tilde{\mathcal{D}}_{\alpha_{\tilde{k}}}^w$ vary in dimension depending on the current mode $\alpha_{\tilde{k}}$.
Therefore, the size of the LMI \eqref{eq:lmiAnalysis-lifted} increases, but the number of decision variables does not, because the dimension of $\tilde{\mathcal{A}}_{\alpha_{\tilde{k}}}^\mathrm{cl}$ is not changing.
The state dimensions of $\tilde{x}$ and $x$ are equal as well.
As mentioned in \cite{Linsenmayer2017}, the lifted graph is typically much smaller, \cf Fig.~\ref{fig:example-WHRTgraph}.
Hence, the overall computational demand for the analysis of the lifted CSS with Theorem~\ref{thm:analysis-lifted} is typically less than for the non-lifted CSS with Theorem~\ref{thm:analysis-nonLifted}, although the dimension of the LMIs increase.
A comparison by means of an example is given in Section~\ref{ch:example}.

%% file: fig/exampleWHRTgraph.tex
\begin{tikzpicture}
\tikzset{vertex/.style = {shape=circle,draw,minimum size=1.7em,inner sep=0}}
\tikzset{vertexFill/.style = {shape=circle,fill=gray!50!white,draw,minimum size=1.6em,inner sep=0}}
\tikzset{edge/.style = {->,> = latex'}}

\begin{scope}[shift={(0,0)}]
	\node[vertexFill] (v1) at (0,0) {$v_1$};
	\node[vertexFill] (v2) at (-0.75,-1) {$v_2$};
	\node[vertexFill] (v3) at (0.75,-1) {$v_3$};
	\node[vertex] (v4) at (-1.1,0.1) {$v_4$};
	\node[vertex] (v5) at (1.1,0.7) {$v_5$};
	\node[vertex] (v6) at (1.75,-0.3) {$v_6$};
	\node[vertex] (v7) at (-2,-1) {$v_7$};

	\draw[edge] (v1) to[loop above] node[above, pos=0.5] {\footnotesize$1$} (v1);
	\draw[edge] (v1) to[bend right] node[above, pos=0.5] {\footnotesize$0$} (v4);
	\draw[edge] (v4) to[bend right] node[right, pos=0.3] {\footnotesize$1$} (v2);
	\draw[edge] (v2) to[bend right] node[left, pos=0.7] {\footnotesize$1$} (v1);
	\draw[edge] (v2) to[bend right] node[above, pos=0.6] {\footnotesize$0$} (v7);
	\draw[edge] (v7) to[bend right] node[below, pos=0.5] {\footnotesize$1$} (v2);
	\draw[edge] (v1) to[bend left] node[above, pos=0.5] {\footnotesize$0$} (v5);
	\draw[edge] (v5) to[bend left] node[above, pos=0.5] {\footnotesize$0$} (v6);
	\draw[edge] (v6) to[bend left] node[below, pos=0.3] {\footnotesize$1$} (v3);
	\draw[edge] (v3) to[bend left] node[below, pos=0.5] {\footnotesize$1$} (v1);

	\node (WHRT23) at ($(0,0)+(0,-1.9)$) {$\mathcal{G}$ for $\lambda = \anyWHRTC{2}{4}$};
\end{scope}

\begin{scope}[shift={(4,0)}]
	\node[vertexFill] (v1) at (0,0) {$v_1$};
	\node[vertexFill] (v2) at (-0.75,-1) {$v_2$};
	\node[vertexFill] (v3) at (0.75,-1) {$v_3$};

	\draw[edge] (v1) to[loop above] node[above, pos=0.5] {\footnotesize$0$} (v1);
	\draw[edge] (v1) to[bend right] node[above left, pos=0.5] {\footnotesize$1$} (v2);
	\draw[edge] (v2) to[bend right] node[left, pos=0.7] {\footnotesize$0$} (v1);
	\draw[edge] (v2) to[loop left] node[above, pos=0.5] {\footnotesize$1$} (v1);
	\draw[edge] (v1) to[bend left] node[above, pos=0.5] {\footnotesize$2$} (v3);
	\draw[edge] (v3) to[bend left] node[below, pos=0.5] {\footnotesize$0$} (v1);

	\node (WHRT23) at ($(0,0)+(0,-1.9)$) {$\tilde{\mathcal{G}}$ for $\lambda = \anyWHRTC{2}{4}$};
\end{scope}

\end{tikzpicture}

%% file: sections/synthesis.tex
\section{CONTROLLER SYNTHESIS} \label{ch:synthesis}
Based on the lifted CSS, we are able to state sufficient LMI conditions which allow to synthesize a state-feedback controller such that the WHRT control system is not only asymptotically stable \cite{Linsenmayer2017}, but additionally has $\ell_2$-performance with gain $\gamma$.
The reason is that in \eqref{eq:switchedSystemClMatrices-lifted} the controller enters linearly, because at each time instant $\tilde{k}$ the control attempt is successful and thus $K$ influences the dynamics at these time instants only.
For all inter-attempt time instants the controller does not appear, because the control signal is lost.
Hence, the controller can be isolated in the LMI \eqref{eq:lmiAnalysis-lifted}, which enables controller synthesis.
The following result aims at synthesizing a non-switched state-feedback controller.
\begin{theorem}[$\ell_2$-performance controller synthesis] \label{thm:synthesis}
	There exists a state-feedback controller such that the lifted CSS \eqref{eq:switchedSystem-lifted}, whose switching is captured by a graph $\tilde{\mathcal{G}}$ according to Definition~\ref{def:WHRTgraph}, has $\ell_2$-performance with gain $\gamma$ if there exist symmetric matrices $S_i \in \mathbb{R}^{n \times n}$, $i = 1,...,\tilde{n}_\mathcal{V}$ and matrices $G \in \mathbb{R}^{n \times n}$, $R \in \mathbb{R}^{m \times n}$ such that
	\begin{align}
		\begin{bmatrix}
			G + G^\top - S_i & \ast & \ast & \ast \\
			0 & \gamma I & \ast & \ast \\
			\tilde{\mathcal{A}}_l G + \tilde{\mathcal{B}}_l R & \tilde{\mathcal{B}}_l^w & S_j & \ast \\
			\tilde{\mathcal{C}}_l G + \tilde{\mathcal{D}}_l R & \tilde{\mathcal{D}}_l^w & 0 & \gamma I
		\end{bmatrix} \succ 0 \quad \forall (i,j,l) \in \mathcal{E}. \label{eq:lmiSynthesis} \raisetag{1\baselineskip}
	\end{align}
	The corresponding controller is then given by $K = RG^{-1}$ and a Lyapunov function is \eqref{eq:lyapFcn-lifted}.
\end{theorem}
\begin{proof}
	Along the lines of \cite{Daafouz2002a}: $G$ is invertible by standard arguments and insert $R = KG$ in \eqref{eq:lmiSynthesis} and together with \eqref{eq:switchedSystemClMatrices-lifted} the conditions of Theorem~\ref{thm:analysis-lifted} are obtained.
\end{proof}
Theorem~\ref{thm:synthesis} can be used in combination with Corollary~\ref{thm:analysis-WHRTsystem} to synthesize a state-feedback controller for the WHRT control system with guaranteed $\ell_2$-performance, while in previous works it was only possible to synthesize a stabilizing controller \cite{Linsenmayer2017}.

As already annotated, Theorem~\ref{thm:synthesis} synthesizes a non-switched controller, which is independent of the past loss sequence.
This induces some conservatism, but is a crucial property for NCS in which the network features no acknowledgment mechanism.
A switched controller requires the knowledge of the past packet losses, which is challenging to achieve without the presence of acknowledgments.
However, our results can directly be extended to switched controllers to be applied to networks that feature such a mechanism by revising the proof by substituting $G$ with $G_i$ and $R$ with $R_i$ in \eqref{eq:lmiSynthesis} with the controller $K_i = R_i G_i^{-1}$.
Switched controllers can improve the feasibility of the control design.
In this case, the control input \eqref{eq:controlInput} and the controller in \eqref{eq:switchedSystemClMatrices-lifted} have to be adapted accordingly to account for the switching.
The resulting controller is dependent on the starting node of the corresponding edge in the lifted WHRT graph, \ie implicitly depends on the past loss sequence and may switch whenever a control attempt is successful.
In \cite{Linsenmayer2021a}, a possibility for designing switched controller policies in the absence of acknowledgment mechanisms is presented, with the tradeoff of requiring computational capabilities at the actuator and having to compute and send larger control input packets.
This technique can be applied here similarly.

Note finally that if a certain performance gain $\gamma$ is guaranteed for a specific $\lambda$, it is also guaranteed for all WHRT control systems with a harder WHRT constraint.

%% file: sections/example.tex
\section{NUMERICAL EXAMPLE} \label{ch:example}
We consider the numerical example from \cite{Blind2015,Linsenmayer2017}, but extended by an performance in- and output, namely
\begin{align}
	\begin{split} \label{eq:exampleSys}
		x_{k+1} &= \begin{bmatrix}
			0 & 1 \\ 1 & 1
		\end{bmatrix} x_k + \begin{bmatrix}
			1 \\ 1
		\end{bmatrix} u_k^\mathrm{a} + \begin{bmatrix}
			1 \\ 1
		\end{bmatrix} w_k\\
		z_k &= \begin{bmatrix}
			1 & 1
		\end{bmatrix} x_k + u_k^\mathrm{a} + w_k.
	\end{split}
\end{align}
The setup is as described in Section~\ref{ch:setup}, \ie the control input losses are described by a WHRT constraint and we use the state-feedback controller \eqref{eq:controlInput}.
We use YALMIP (Version R20210331) \cite{Lofberg2004} and MOSEK \cite{mosek} to minimize $\gamma$ with the respective LMIs as constraints.
Note that the obtained $\gamma$ is in general only an upper bound on the $\ell_2$-gain of the WHRT control system, since the presented conditions are only sufficient.
In \cite{Blind2015}, the controller $K = \begin{bmatrix}-0.35 & -0.85\end{bmatrix}$ for the WHRT constraint $\lambda = \anyWHRTC{2}{3}$ and the zero strategy was proposed.
For this controller, Theorems~\ref{thm:analysis-nonLifted} and~\ref{thm:analysis-lifted} yield $\gamma = 3.52$.
Simulating this system with $w_k = 1$ for $k \geq T$ and $w_k = 0$ otherwise with sufficiently large, finite $T$ and the worst-case loss sequence $\mu=(1\,0\,1\,1\,0\,1...)$, the simulated $\gamma$ obtained by computing \eqref{eq:defPerformance} is $\gamma_\mathrm{sim} = 3.38$.
Note that it is in general difficult to find the worst-case input $w$ and loss sequence $\mu$ that lead to the highest $\gamma$, \ie the $\ell_2$-gain.
By synthesizing a non-switched controller with Theorem~\ref{thm:synthesis} we obtain the controller $K = \begin{bmatrix} -0.61 & -1.00 \end{bmatrix}$, which results in $\gamma = 2.505$.
The respective simulated gain is $\gamma_{\mathrm{sim}} = 2.10$.
Therefore, we are able to significantly improve the $\ell_2$-performance of \eqref{eq:exampleSys}.
If we synthesize a switched controller, the $\ell_2$-performance can be improved even a bit further to $\gamma = 2.488$.
Note that the improvement of a switched controller over a non-switched one strongly depends on the chosen system and the WHRT constraint.

To compare the computational complexity, we consider the WHRT constraint $\anyWHRTC{4}{10}$.
The corresponding non-lifted WHRT graph has $462$ edges and $336$ nodes and its lifted counterpart has $210$ edges and $84$ nodes.
This leads to $462$ LMIs with $\num{2352}$ decision variables, or $210$ LMIs with $\num{588}$ decision variables.
This significant reduction can also be seen in the computation time.
For above example, the analysis with Theorem~\ref{thm:analysis-nonLifted} took around \SI{1.1}{\second} to compute on a standard computer, while for the lifted system it finished after roughly \SI{0.3}{\second}.
Thus, our numerical findings confirm the computational advantages when considering the lifted system instead of the non-lifted one.

%% file: sections/conclusion.tex
\section{CONCLUSIONS} \label{ch:conclusion}
In this paper, we considered the $\ell_2$-performance of linear plants with unreliable feedback, \ie the control signal can be lost, which is modeled by a WHRT constraint.
We presented sufficient LMI conditions under which such systems have $\ell_2$-performance with a certain gain $\gamma$.
Moreover, we proposed an approach for the synthesis of state-feedback controller that ensures that the WHRT control system has $\ell_2$-performance with gain $\gamma$.
This was made possible by lifting the system, which in addition typically reduces the computational complexity of the respective LMI conditions for the $\ell_2$-performance analysis.
The resulting controller can be a non-switched, loss sequence independent one for NCS without acknowledgment mechanisms, or designed as a switched controller, that switches depending on the past loss sequence.
As a side result, we also obtained an $\ell_2$-performance analysis result for general switched systems with constrained switching that can be captured by a graph.

The presented results can be generalized to a dissipativity framework, allowing for robust quadratic performance of CSS and WHRT control systems, see \cite{Lang2024}.

To make the results more applicable to NCS and real-time control systems, a potential future line of research is the extension to nonlinear plants.
Another interesting open problem is to consider output feedback controllers.
Further research can also be conducted the direction of how an actuator with computational capabilities can improve the control performance.